\documentclass[11pt]{article}
\usepackage{coling2016}
\usepackage{times}
\usepackage{url}
\usepackage{latexsym}

\usepackage{graphicx}
\usepackage[english]{babel}
\usepackage{subfigure}
\usepackage{multirow}
\usepackage{threeparttable}
\usepackage{CJK}
\usepackage{mathrsfs}
\usepackage{amssymb}
\usepackage{amsmath}
\usepackage{algorithm}
\usepackage{algorithmic}

\title{Hierarchical Memory Networks for Answer Selection on Unknown Words}

\author{Jiaming Xu$^{a,*}$, Jing Shi$^{a,*}$, Yiqun Yao$^{a}$, Suncong Zheng$^{a}$, Bo Xu$^{a}$, Bo Xu$^{a,b}$\\
  $^a$Institute of Automation, Chinese Academy of Sciences (CAS). Beijing, China\\
  $^b$Center for Excellence in Brain Science and Intelligence Technology, CAS. China \\
  {\tt \{jiaming.xu, shijing2014, yaoyiqun2014\}@ia.ac.cn} \\
  {\tt \{suncong.zheng, boxu, xubo\}@ia.ac.cn}}

\date{}

\begin{document}
\maketitle
\begin{abstract}
Recently, end-to-end memory networks have shown promising results on Question Answering task, which encode the past facts into an explicit memory and perform reasoning ability by making multiple computational steps on the memory. However, memory networks conduct the reasoning on sentence-level memory to output coarse semantic vectors and do not further take any attention mechanism to focus on words, which may lead to the model lose some detail information, especially when the answers are rare or unknown words. In this paper, we propose a novel Hierarchical Memory Networks, dubbed HMN. First, we encode the past facts into sentence-level memory and word-level memory respectively. Then, \(k\)-max pooling is exploited following reasoning module on the sentence-level memory to sample the \(k\) most relevant sentences to a question and feed these sentences into attention mechanism on the word-level memory to focus the words in the selected sentences. Finally, the prediction is jointly learned over the outputs of the sentence-level reasoning module and the word-level attention mechanism. The experimental results demonstrate that our approach successfully conducts answer selection on unknown words and achieves a better performance than memory networks.
\end{abstract}

\section{Introduction}
\label{intro}

%
%

\blfootnote{
    \hspace{-0.65cm}  
    This work is licensed under a Creative Commons
    Attribution 4.0 International Licence.
    Licence details:
    \url{http://creativecommons.org/licenses/by/4.0/}
    
    $^{*}$The first two authors contributed equally.
}

With the recent resurgence of interest in Deep Neural Networks (DNN), many researchers have concentrated on using deep learning to solve natural language processing (NLP) tasks~\cite{19_collobert2011natural,9_sutskever2014sequence,29_zeng2014relation,16_feng2015applying}. The main merits of these representation learning based methods are that they do not rely on any linguistic tools and can be applied to different languages or domains. However, the memory of these methods, such as Long Short-Term Memory (LSTM)~\cite{21_hochreiter1997long} and Gated Recurrent Unit (GRU)~\cite{20_cho2014learning} compressing all the external sentences into a fixed-length vector, is typically too small to accurately remember facts from the past, and may lose important details for response generation~\cite{27_shang2015neural}. Due to the drawback, these traditional DNN models encounter great limitation on Question Answering (QA), as a complex NLP task, which requires deep understanding of semantic abstraction and reasoning over facts that are relevant to a question~\cite{1_hermann2015teaching,14_yu2015empirical}.

Recently, lots of deep learning methods with explicit memory and attention mechanism are explored for Question Answering (QA) task, such as Memory Networks (MemNN)~\cite{5_sukhbaatar2015end}, Neural Machine Translation (NMT) and Neural Turing Machine (NTM)~\cite{14_yu2015empirical}. These methods exploit a external memory to store the past sentences with a continuous representation and utilize attention mechanism to automatically soft-search for parts of the memory for prediction. Compared with NMT and NTM, MemNN, making multiple computational steps (termed as ``hops'') on the memory before making an output, is better qualified for textual reasoning tasks. However, for QA task, MemNN only conducts the reasoning on sentence-level memory and does not further take any attention mechanism to focus on words in the retrieved facts. More recently, \newcite{14_yu2015empirical} constructed a {\emph{Search-Response}} pipeline where {\emph{Search}} component uses MemNN to search the supporting sentences and {\emph{Response}} component uses NMT or NTM to generate answer on the selected sentences. However, that work needs the supervision of the supporting facts to guide the training of {\emph{Search}} component and the combination of these two components through a separate training way may hurt the performance. Along the direction of that work, we believe that a joint learning model can achieve a better performance by designing a hierarchical architecture, with sentence-level and word-level components, which has shown promising results on document modeling~\cite{22_lin2015hierarchical} and document classification~\cite{23_yang2016hierarchical}.

Besides, rare and unknown word problem as an important issue should be considered in NLP tasks, especially for QA task, where the words that we are mainly interested in are usually named entities which are mostly unknown or rare words~\cite{28_marrero2013named,18_gulcehre2016pointing}. In order to control the computational complexity, many methods limit the trained vocabulary size, which further leads to lots of low-frequency words outside the trained vocabulary~\cite{17_li2016towards}. Traditional methods directly mask the rare or unknown words with meaningless {\emph{unk}} which may lose the important information for answer selection task. For example, given a set of sentences as follows:

~~~~1.	Miss, what is your name?

~~~~2.	Uh, my name is {\emph{Wainwright}}.

~~~~3.	Please tell me your passport number.

~~~~4.	Ok, it is {\emph{899917359}}.

Assume that the words {\emph{Wainwright}} and {\emph{899917359}}\footnote{Note that the personal information used in our examples and datasets throughout this paper is all synthetic and not real.} are rare words or outside the trained vocabulary. If these words are discarded or replaced with {\emph{unk}} symbol, any models may not be able to select the correct answers for response during testing.

Based on the above observations, this paper proposes a Hierarchical Memory Networks\footnote{It is worth noticing that the term ``Hierarchical Memory Networks'' has been mentioned in~\cite{25_chandar2016hierarchical} where the intention was to organize the memory into multi-level groups based on hashing, tree or clustering structures to make the reader efficiently access the memory, whereas in our paper the term has a different meaning.} (dubbed to HMN) for answer selection. Our method first maps the sentences into a sentence-level memory and reasoning module takes multiple hops on the sentence-level memory to soft-search the related sentences. Meanwhile, all words in the sentences are encoded into a word-level memory with recurrent neural networks. Then, we exploit \(k\)-max pooling to sample the most relevant sentences and feed these selected sentences into attention mechanism on the word-level memory to focus the words. Finally, the prediction is jointly learned over the outputs of the sentence-level reasoning module and the word-level attention mechanism. Our main contributions are three-fold:

(1). We proposed a novel hierarchical memory networks for answer selection, where the reasoning module is performed on sentence-level memory to retrieve the relevant sentences and the attention mechanism is applied on word-level memory to focus the words. This hierarchical architecture allows the model to have explicit reasoning ability on sentences and also focus on more fine-grained words.

(2). \(k\)-max pooling is exploited to sample the most relevant facts based on the results of the sentence-level reasoning and then feed these facts into word-level attention mechanism, which can filter the noise information and also reduce the computational complexity on word-level attention.

(3). We release four synthetic domain dialogue datasets\footnote{Our code and dataset are available:~\url{https://github.com/jacoxu/HMN4QA}}, two from air-ticket booking domain and two from hotel reservation domain, where the answers are mostly rare or unknown words, and lots of answers should be reasoned based on some supporting sentences. The experimental results show that our approach can successfully conduct answer selection on unknown words.

\section{Background: Memory Networks}
\label{sec:BackgroundMemoryNetworks}

Here, we give a brief description of memory networks which have shown promising results on QA tasks~\cite{4_weston2014memory,3_bordes2015large,5_sukhbaatar2015end}. Memory network first introduced by \newcite{4_weston2014memory} is a new class of learning models which can easily read and write to part of a long-term memory component, and combine this seamlessly with inference for prediction. Formally, besides the explicit memory which is an array of cells to memorize the pre-trained vector representations of the external data, a general memory network consists of four major components: (1). {\emph{Input feature map}} which converts the incoming input to the internal feature representation. (2). {\emph{Generalization}} which updates old memories given the new input. (3). {\emph{Output feature map}} which produces a new output based on the new input and the current state. (4). {\emph{Response}} which converts the output into the response format desired. Along the above framework, \newcite{5_sukhbaatar2015end} put forward end-to-end memory networks which do not require the supervision of the supporting facts and are more generally applicable in realistic setting. Thus, we choose end-to-end memory networks, denoted as MemNN throughout our paper, as the foundation of our proposed approach.

\section{Hierarchical Memory Networks for Answer Selection}
\label{sec:HierarchicalMemoryNetworks}

\subsection{Approach Overview}
\label{ssec:ApproachOverview}

\begin{figure*}[t]
\begin{center}
\includegraphics[width=16cm]{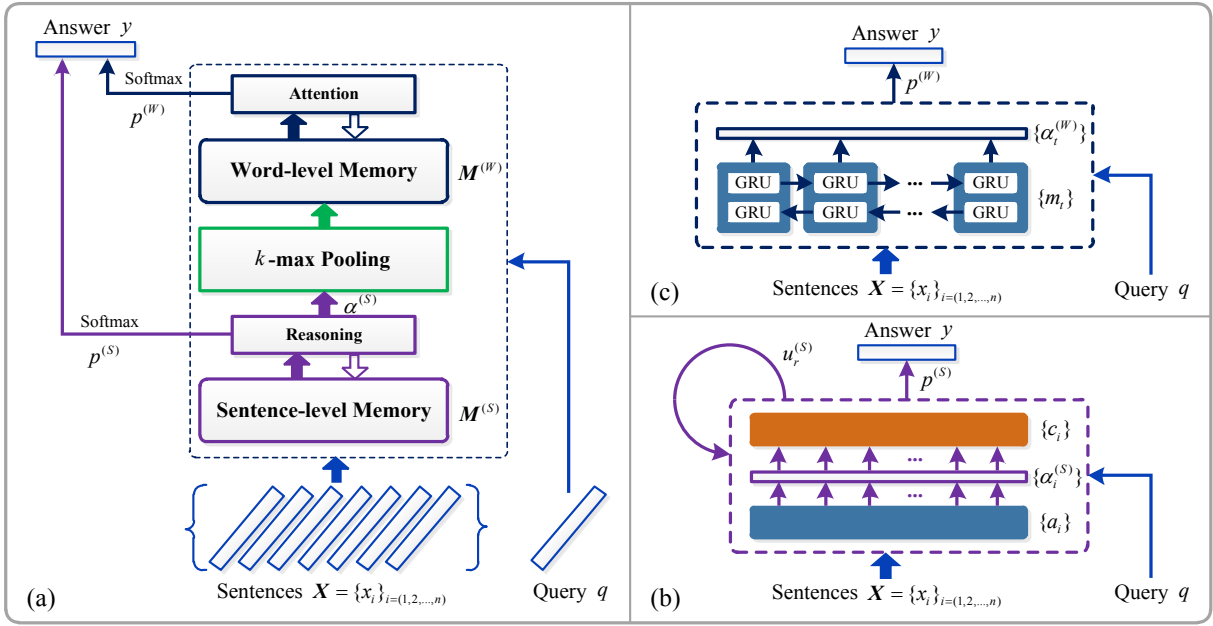}
\caption{An illustration of our Hierarchical Memory Networks (HMN). (a): The overall architecture of the proposed HMN. (b): Reasoning module of our approach on sentence-level memory. (c): Attention module of our approach on word-level memory.}\label{fig:MethodOverview}
\end{center}
\end{figure*}

As described in Figure~\ref{fig:MethodOverview}(a), we give an illustration of our HMN for answer selection. Given a set of \(n\) sentences denoted as: \({\bf{X}} = {\{ {x_i}\} _{i = (1,2,...,n)}}\) and a query \(q\), where \(i\) is the timestep of sentence \(x_i\) in the set. We first map these sentences \({\bf{X}}\) into the sentence-level memory \({{\bf{M}}^{(S)}}\) and the word-level memory \({{\bf{M}}^{(W)}}\) with low-dimensional distributed representations respectively. Then, reasoning on the sentence-level memory is utilized to soft-search the related sentences. We further exploit \(k\)-max pooling to sample the most relevant sentences based on the soft-searching results and take attention mechanism to focus on word-level memory of the selected sentences. The target answer \(y\) is used to guide the learning of the reasoning on sentence-level memory and the attention on word-level memory learning simultaneously.

\subsection{Sentence-level Memory and Reasoning}
\label{ssec:SentLevelMemory}

In this section, we apply reasoning module to make multiple interaction on sentence-level memory based on the adjacent weight tying scheme of MemNN~\cite{5_sukhbaatar2015end}, as shown in Figure~\ref{fig:MethodOverview}(b). Given two word embedding matrices \({\bf{A}} \in {\mathbb{R}^{|V| \times d}}\) and \({\bf{C}} \in {\mathbb{R}^{|V| \times d}}\), where \(|V|\) is the vocabulary size and \(d\) is the dimension of the word embedding, we first encode the word \({x_{ij}}\) at timestep \(j\) in the sentence \({x_i}\) into dual channels of word representation as \({\bf{A}}{x_{ij}} \in {\mathbb{R}^d}\) and  \({\bf{C}}{x_{ij}} \in {\mathbb{R}^d}\).

In order to combine the order of the words into their representations, a positional encoding matrix \({\bf{l}}\) is applied to update the dual-channel word embeddings as \({l_{gj}} \cdot ({\bf{A}}{x_{ij}})\) and \({l_{gj}} \cdot ({\bf{C}}{x_{ij}})\), where

\begin{equation}
{l_{gj}} = (1 - j/{J_i}) - (g/d)(1 - 2j/{J_i}),~~~~1 \le j \le {J_i},~1 \le g \le d,
\label{eq:UpdateMatrixL}
\end{equation}
and \({J_i}\) is the length of the sentence \({x_i}\) and \(g\) is the embedding index. This positional encoding scheme is also successfully applied in~\cite{26_xiong2016dynamic}.

Two temporal encoding matrices \({{\bf{T}}_A} \in {\mathbb{R}^{n \times d}}\) and \({{\bf{T}}_C} \in {\mathbb{R}^{n \times d}}\) are further utilized to encode the order of the sentences. Then, the sentence-level memory \({{\bf{M}}^{(S)}} = \{ \{ {a_i}\} ,{\rm{ }}\{ {c_i}\} \}_{i = (1,2,...,n)} \) is reformed as:

\begin{equation}
{a_i} = \sum\nolimits_j {{l_j} \cdot ({\bf{A}}{x_{ij}}) + {{\bf{T}}_A}(i)},~~~~{c_i} = \sum\nolimits_j {{l_j} \cdot ({\bf{C}}{x_{ij}}) + {{\bf{T}}_C}(i)},
\label{eq:MemoryOfSent}
\end{equation}
where \({l_j}\) is the \(j\)-th column vector of the position encoding matrix \({\bf{l}}\) according to the sentence \({x_i}\) and the operation ``\(\cdot\)'' means the element-wise multiplication.

For the query \(q\), the \(j\)-th word \({q_j}\) is also embedded as \({\bf{A}}{q_j} \in {\mathbb{R}^d}\), where the \({\bf{A}}\) is the embedding matrix used in Eqn.~(\ref{eq:MemoryOfSent}). By encoding the word position \(j\) into the query representation, we get the probe representation of the query \(q\) as follows:
\begin{equation}
u_1^{(S)} = \sum\nolimits_j {{l_j} \cdot ({\bf{A}}{q_j})},
\label{eq:QueryUpdate}
\end{equation}
where \({l_j}\) is the \(j\)-th column vector of the position encoding matrix \({\bf{l}}\) according to the query \(q\). Then the attention weights of the sentences according to the query can be calculated through the inner product of the two vectors as \(\alpha _i^{(S)} = softmax (a_i^Tu_1^{(S)})\), and the output of the sentence-level memory based on the activation of the query can be obtained as: \({o_1} = \sum\nolimits_i {\alpha _i^{(S)}{c_i}} \).

In order to perform reasoning on sentence-level memory to find the most relevant sentences, we make \(R\) hops to soft-search the sentences and output the final vector \({o_R}\). To be specific, during the \(r + 1\) hop of the reasoning operation, the process can be formalized as: \(u_{r + 1}^{(S)} = {o_r} + u_r^{(S)}\), \(\alpha _i^{(S)} = softmax (a_i^Tu_{r + 1}^{(S)})\), \({o_{r + 1}} = \sum\nolimits_i {\alpha _i^{(S)}{c_i}} \), and the dual-channel memories are updated as follows:

\begin{equation}
{a_i} = \sum\nolimits_j {{l_j} \cdot ({{\bf{A}}^{r + 1}}{x_{ij}}) + {\bf{T}}_A^{r + 1}(i)},~~~~{c_i} = \sum\nolimits_j {{l_j} \cdot ({{\bf{C}}^{r + 1}}{x_{ij}}) + {\bf{T}}_C^{r + 1}(i)},
\label{eq:MemoryOfSentR}
\end{equation}
where \(1 \le r \le (R - 1)\). Specifically, during the \(r + 1\) hop, the word embedding matrices \({{\bf{A}}^{r + 1}}\) and \({{\bf{C}}^{r + 1}}\) are mutually independent, so as the temporal encoding matrices \({\bf{T}}_A^{r + 1}\) and \({\bf{T}}_C^{r + 1}\). But during the adjacent two hops, \({{\bf{A}}^{r + 1}} = {{\bf{C}}^r}\) and \({\bf{T}}_A^{r + 1} = {\bf{T}}_C^r\). Finally, we can get the predicted word probability distribution by applying softmax on the output vector of the reasoning on the sentence-level memory as:

\begin{equation}
{p^{(S)}}(w) = softmax ({({{\bf{C}}^R})^T}({o_R} + u_R^{(S)})),
\label{eq:AttentionOfSentR}
\end{equation}
where \(w = {\{ {w_t}\} _{t = (1,2,...,|V|)}}\) is the word set with a vocabulary size of \(\left| V \right|\), the weight matrix is the same as the embedding matrix \({{\bf{C}}^R} \in {\mathbb{R}^{|V| \times d}}\) on the last hop, and \(T\) is the operation of matrix transposition.

\subsection{${k}$-max Pooling}
\label{ssec:KmaxPooling}

Here, we exploit a pooling operation over the top attention weights \(\alpha^{(S)}\) of the reasoning module on the sentence-level memory to sample the most relevant sentences. Given a value \(k\) and the top attention weights \(\alpha^{(S)}\) of length \(n \ge k\), we use the \(k\)-max pooling to select a subset of sentence sequences \(\hat {\bf{X}} = {\{ {\hat x_i}\} _{i = (1,2,...,k)}}\), corresponds with their top-\(k\) maximum values of \({\alpha ^{(S)}}\) on the sentences.
	
The \(k\)-max pooling operation makes it possible to pool the \(k\) most relevant sentences to the query and filter the noise information, which maybe more beneficial to select the correct answers. Moreover, this sampling module feeds a subset of sentences \(\hat {\bf{X}}\) to the following attention mechanism on the word-level memory, which can reduce the computation complexity of the attention to focus on the relevant words.

\subsection{Attention on Word-level Memory}
\label{ssec:WordLevelMemory}

For word-level memory, we first apply a Bi-directional GRU (BiGRU) to compute the hidden states of all the ordered words \(\bar w = {\{ {\bar w_t}\} _{t = (1,2,...,|t|)}}\) in the sentence set \({\bf{X}}\), where \(|t|\) is the time steps of the words in the sentences. In particular, for the \(t\)-th word \({\bar w_t}\), the forward GRU and the backward GRU encode it as hidden states \({\vec h_t} = \mathop {GRU}\limits^ \to  ({{\bf{C}}^R}{\bar w_t})\) and \({\mathord{\buildrel{\lower3pt\hbox{$\scriptscriptstyle\leftarrow$}}
\over h} _t} = \mathop {GRU}\limits^ \leftarrow  ({{\bf{C}}^R}{\bar w_t})\) respectively, where \({{\bf{C}}^R}\) is the word embedding matrix of the last hop on the sentence-level memory, and we set the dimension of \({\vec h_t}\) and \({\mathord{\buildrel{\lower3pt\hbox{$\scriptscriptstyle\leftarrow$}}
\over h} _t}\) equals to the dimension of the word embedding. By summing the forward hidden states and the backward hidden states, we obtain the word-level memory as \({{\bf{M}}^{(W)}} = {\{ {m_t}\} _{t = (1,2,...,|t|)}}\), where \({m_t} = {\vec h_t} + {\mathord{\buildrel{\lower3pt\hbox{$\scriptscriptstyle\leftarrow$}}
\over h} _t}\). In this way, the memory \({m_t}\) contains the context information of the \(t\)-th word \({\bar w_t}\) in the sentence set \({\bf{X}}\).

Then, we perform attention on the subset of the ordered words \({\hat w}={\{ {\hat w_t}\} _{t = (1,2,...,|\hat t|)}}\) in the selected sentences \({\bf{\hat X}}\) by using the probe vector \(u_R^{(S)}\) of the last hop on the sentence-level memory and a subset of word-level memory \({\{ {\hat m_t}\} _{t = (1,2,...,|\hat t|)}}\) selected form \({{\bf{M}}^{(W)}}\) according to the word subset \(\hat w\). The normalized attention weights \(\alpha _{}^{(W)} = {\{ \alpha _t^{(W)}\} _{t = (1,2,...,|\hat t|)}}\) on the word-level memory are calculated as:

\begin{equation}
\alpha _t^{(W)} = softmax ({v^T}tanh({\bf{W}}u_R^{(S)} + {\bf{U}}{\hat m_t})),
\label{eq:AttentionOfWord}
\end{equation}
where \(v \in {\mathbb{R}^{d \times 1}}\), \({\bf{W}} \in {\mathbb{R}^{d \times d}}\) and \({\bf{U}} \in {\mathbb{R}^{d \times d}}\) are all learning parameters updated during the training. Inspired by Pointer Networks~\cite{12_vinyals2015pointer}, we adopt the normalized attention weights \(\alpha _{}^{(W)}\) on the word collection \(\hat w\) as the probability distribution of the output words:

\begin{equation}
{p^{(W)}}(w) = trans({p^{(W)}}(\hat w)) = trans({\alpha ^{(W)}}),
\label{eq:PointOutput}
\end{equation}
where \(trans( \cdot )\) means the operation to map the words probability distribution \({p^{(W)}}(\hat w) \in {\mathbb{R}^{|\hat t|}}\) into the probability distribution \({p^{(W)}}(w) \in {\mathbb{R}^{|V|}}\). To be specific, the map operation makes the probability distribution \({p^{(W)}}(\hat w)\) of the word subset (\(\hat w = {\{ {\hat w_t}\} _{t = (1,2,...,|\hat t|)}}\)) to be added into their corresponding positions in the vocabulary (\(w = {\{ {w_t}\} _{t = (1,2,...,|V|)}}\)), and the probabilities of the words not in selected word subset \({\hat w}\) will be set to zero\footnote{For example, if the word ``airport'', at two different timesteps in the ordered word subset \(\hat w\), has two probabilities ``0.1'' and ``0.3'', the map operation would add up these probabilities and set ``0.4'' as the probability of the word ``it'' in the vocabulary.}. Finally, we get the new probability distribution \({p^{(W)}}(w) \in {\mathbb{R}^{|V|}}\).

\begin{table}[t]
\begin{center}
\begin{tabular}{|l|c|c|c|c|}
\hline Domain~(Lang) &     Train/Dev/Test &     Vocab &     Unseen Answers~(Dev/Test) &     Max Len~(P/S) \\ \hline
{\emph{Air-Ticket~(CH)}}&           5,400/600/6,000 &    8,540 &     409~(68.2\%)/4,020~(67.0\%) &        16/17     \\
{\emph{Hotel~(CH)}} &	           5,400/600/6,000	&   7,586 &     367~(61.2\%)/3,690~(61.5\%) &        16/16     \\
{\emph{Air-Ticket~(EN)}} &	       5,400/600/6,000	&   7,537 &     342~(57.0\%)/3,489~(58.2\%) &        16/18     \\
{\emph{Hotel~(EN)}} &	           5,400/600/6,000	&   7,134 &     357~(59.5\%)/3,452~(57.5\%) &        16/16     \\
{\emph{Total}} &                 21,600/2,400/24,000	&   29,092&     1,406~(58.6\%)/13,872~(57.8\%) &	 16/18     \\

\hline
\end{tabular}
\end{center}
\caption{\label{tb:DataDescription} Statistics of the datasets, including the domain and the language of the dataset (CH: Chinese and EN: English), the number of train/dev/test set entries, the vocabulary size of datasets, the number and proportion of unseen answers on dev/test set, and the max paragraph length and sentence length of the datasets. The {\emph{Total}} dataset consists all the samples of the above four datasets.}
\end{table}

\subsection{Joint Learning}
\label{ssec:JointLearning}
In this paper, we combine the probability distributions of the output words both on the sentence-level memory and the word-level memory to predict the joint probability distribution \(p(w)\) as follows:

\begin{equation}
p(w) = {p^{(S)}}(w) + {p^{(W)}}(w).
\label{eq:JointLearning}
\end{equation}


Finally, we use the target answer \(y\) to guide the learning of the reasoning module on sentence-level memory and the attention module on word-level memory simultaneously. We choose the cross entropy as the cost function and apply Stochastic Gradient Descent (SGD)~\cite{24_bottou1991stochastic} as the optimization method to train our joint model. The learned parameters include word embedding matrices \( {{\bf{A}}^1}\) and \({\{ {{\bf{C}}^r}\} _{r = (1,2,...,R)}}\), temporal encoding matrices \( {\bf{T}}_A^1\) and \({\{ {\bf{T}}_C^r\} _{r = (1,2,...,R)}}\) in Eqn.~(\ref{eq:MemoryOfSent}) and (\ref{eq:MemoryOfSentR}), the parameters \(\{ {\theta _{BiGRU}}\} \) of the BiGRU model and the attention parameters \(v\), \({\bf{W}}\) and \({\bf{U}}\) in Eqn.~(\ref{eq:AttentionOfWord}).

\section{Experiments}
\label{sec:Experiments}

\begin{table}[t]
\begin{center}
\begin{tabular}{|l|c|c|c|c|c|}
\hline ~~ &     {\emph{Air-Ticket (CH)}} &     {\emph{Hotel (CH)}} &     {\emph{Air-Ticket (EN)}} &     {\emph{Hotel (EN)}} & {\emph{Total}} \\ \hline
MemNN-H1 &  4,727.8$\pm$79.2   &   3,680.8$\pm$48.3   &   4,051.8$\pm$53.8   &   3,067.4$\pm$71.3   &   14,492.8$\pm$282.8   \\
MemNN-NT &	3,424.0$\pm$106.6   &   2,816.2$\pm$35.9   &   2,984.2$\pm$93.0   &   2,304.0$\pm$70.3   &   10,250.4$\pm$152.1  \\
MemNN    &   67.5$\pm$9.1   &   84.0$\pm$9.8   &   55.5$\pm$9.4   &   56.2$\pm$10.7   &   225.6$\pm$41.1   \\ \hline
HMN-Sent &	     99.8$\pm$29.9   &   175.0$\pm$19.3  &   112.8$\pm$51.2	&    129.2$\pm$15.3	&    125.6$\pm$27.4\\
HMN-Word&        22.2$\pm$7.5	& 31.8$\pm$7.8	&  24.8$\pm$9.6	&  12.8$\pm$4.9	&  27.4$\pm$4.8    \\
HMN-Joint &	     {\bf{4.2$\pm$2.5}}	&  {\bf{9.8$\pm$1.9}}	&   {\bf{4.4$\pm$2.7}}	&   {\bf{3.0$\pm$1.2}}	&   {\bf{4.2$\pm$1.8}}    \\

\hline
\end{tabular}
\end{center}
\caption{\label{tb:MethodComparison}Comparison of predicted test error numbers of our HMN and MemNN with different components on four domain datasets (Test: 6,000 samples) and the {\emph{Total}} dataset (Test: 24,000 samples). MemNN-H1: Memory network with 1 hop and temporal encoding, MemNN-NT: Memory network with 3 hops but without temporal encoding. MemNN: Memory network with 3 hops and temporal encoding. Note that HMN-Sent and HMN-Word, as the parts of HMN-Joint, are joint learning but give their predicted answers separately.}
\end{table}

\subsection{Datasets and Setup}
\label{ssec:Datasets}
We conduct answer selection tasks on four synthetic domain dialogue datasets, two from air-ticket booking domain and two from hotel reservation domain. One complete dialogue history of each dataset has eight round responses. Besides greeting and ending sentences, one dialogue history consists six round responses to query and answer client's personal information, such as name, phone and passport number. The datasets contain hundreds of response patterns and thousands of entity information. More detailed descriptions can be found in our released datasets. The statistics of the datasets are summarized in Table~\ref{tb:DataDescription}. We use 45\% of the data for training, 5\% for validation and the remaining 50\% for test. From the statistics, we can see that the proportions of the unseen answers on dev/test sets all overtake 57\%.

In our experiments, the most of hyper parameters are set uniformly for the datasets as described in Table~\ref{tb:Hyperparameter}. The training gradients with an \(l2\) norm larger than 40 are clipped to 40 and the learning rate is annealed every 15 epochs by \(\lambda /2\) until 60 epochs are reached. The learned parameters are all initialized randomly from a Gaussian distribution with zero mean and 0.1 standard deviation. In order to make the comparison more intuitive, we use the number of predicted error samples on each dataset to evaluate the performance for answer selection and calculate the average result by repeating each experiment 5 times.

\begin{table}[th]
\begin{center}
\begin{tabular}{|c|c|c|c|c|c|}
\hline Hyperparameter & Hidden dim. & Hops & Max pooling & Learning rate & Batch size ({\emph{Total}}) \\ \hline
Value & $d = 100$ & $R = 3$ & $k = 4$ & $\lambda  = 0.01$ & 30 (32) \\
\hline
\end{tabular}
\end{center}
\caption{\label{tb:Hyperparameter}Hyperparameters used in our experiments. The dimension of word embeddings and the dimension of GRU hidden states are set equally to 100, batch sizes are 32 for {\emph{Total}} and 30 for the others.}
\end{table}

\subsection{Comparison with Memory Networks}
\label{ssec:ComparisonWithMemoryNetworks}

In order to evaluate the effect of multiple hops for reasoning module and temporal encoding on sentence-level memory, we design three MemNN~\cite{5_sukhbaatar2015end} based variants: MemNN-H1 (1 hop and temporal encoding), MemNN-NT (3 hops but not temporal encoding) and MemNN (3 hops and temporal encoding) on the datasets. We further evaluate the prediction performance of our HMN on different level memory components via HMN-Sent (prediction of reasoning module on sentence-level memory as Eqn.~(\ref{eq:AttentionOfSentR})), HMN-Word (prediction of attention module on word-level memory as Eqn.~(\ref{eq:PointOutput})) and HMN-Joint (joint prediction as Eqn.~(\ref{eq:JointLearning})). The comparison of these methods are reported in Table~\ref{tb:MethodComparison}. From the results, we can see that MemNN-H1 without temporal encoding and MemNN-NT without multi-hop reasoning make the worst performances, which clearly demonstrate that multiple hops for reasoning module and temporal encoding on sentence-level memory play a very important role on our tasks. Despite that MemNN represents surprising results on this task, our HMN-Word and HMN-joint further improve the answer selection performance on all the datasets. Compared with the results of HMN-Sent and HMN-Word, the results also show that the joint prediction can make a better performance.

%

\begin{figure*}[t]
\begin{center}
\includegraphics[width=16cm]{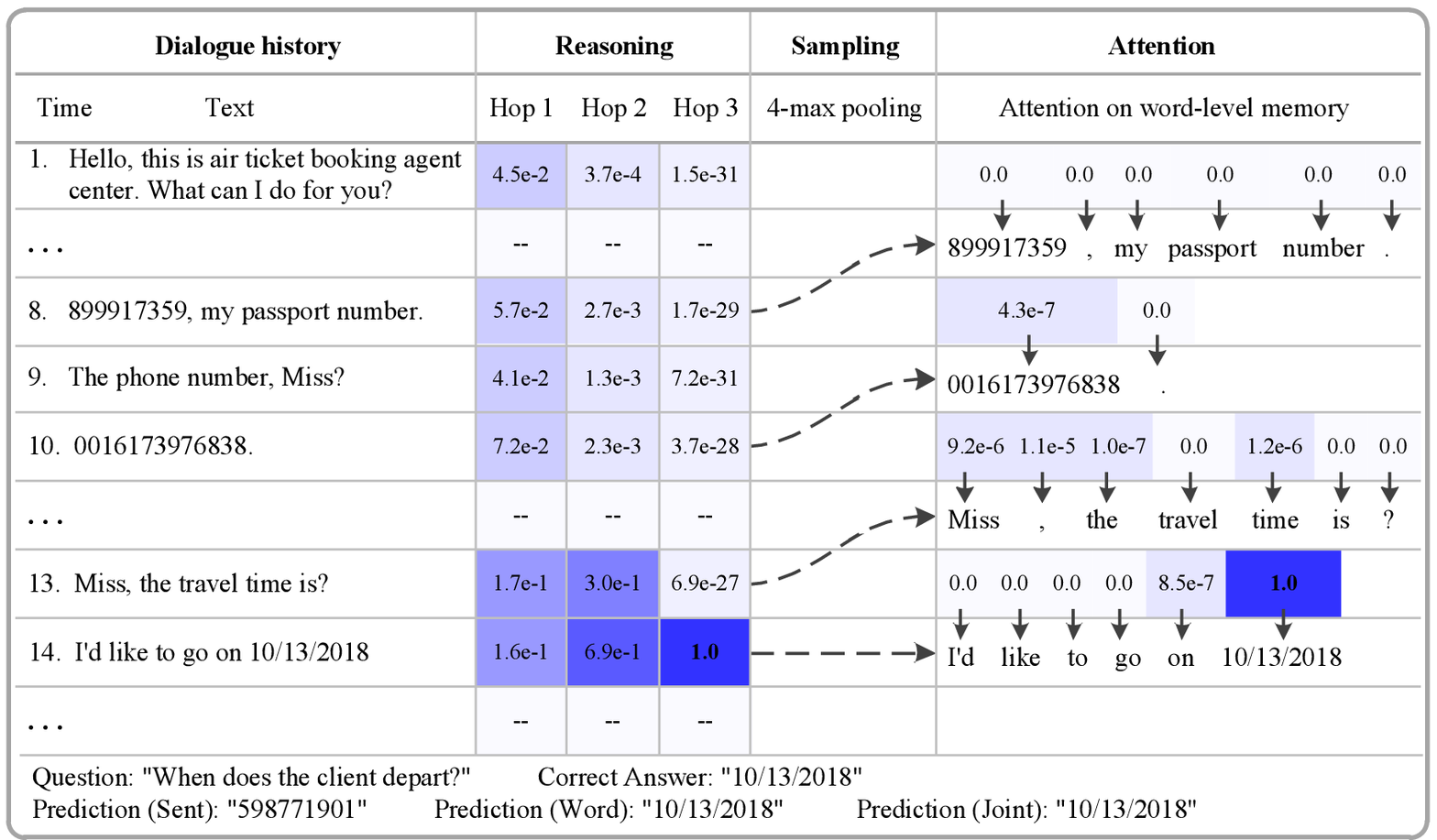}
\caption{\label{fig:AttentionGraph}One example prediction by our HMN for answer selection. The hop columns in reasoning module show the probabilities of each hop, and the sampling module selects the 4-max relevant sentences and feeds these selected sentences to the attention module on word-level memory. In the experiments, all the sentences in the dialogue history are indexed in reverse order to reflect their relative distance.}
\end{center}
\end{figure*}

\subsection{How to Select the Correct Answers on Unknown Words}
\label{ssec:HowToTakeAttentionOnUNK}

Here, we try to answer two questions: (1) How does the reasoning module focus on the related sentences and predict the rare and unknown words on sentence-level memory? (2) How does the attention module focus on the correct answers and distinguish multiple rare and unknown words on word-level memory? We give a visual example of our HMN over one dialogue history for answer selection in Figure~\ref{fig:AttentionGraph} to get a better understanding of the effect of each module. The example is one dialogue history from air-ticket booking domain which contains 16 sentences associated with their temporal indexs.

From Figure~\ref{fig:AttentionGraph}, we can see that in the first hop, the reasoning module mainly focuses on the sentences 13 and 14, which have most semantic relevance to the question ``When does the client depart?''. As the effect of temporal encoding as Eqn.~(\ref{eq:MemoryOfSentR}), the reasoning allocates more weight to the most related sentence 14 in the following hops. Another interesting result as shown in Table~\ref{tb:MethodComparison} is that MemNN and HMN-Sent represent surprising performance to predict the rare and unknown words on sentence-level memory. An explanation maybe that the way in which these methods use a simple way, rather than sophisticated LSTM or GRU, to encoding the sentence into memory as Eqn.~(\ref{eq:UpdateMatrixL}). This simple sentence encoding strategy can remain the raw embedding representation of words, and MemNN and HMN-Sent utilize the transpose of the raw embedding matrix as the decoding weights to conduct answer match as Eqn.~(\ref{eq:AttentionOfSentR}) in the raw embedding space. Nonetheless, sentence-level encoding may introduce other semantic information which may lead to predict an error answer, as the prediction (Sent) in the example.

\begin{figure*}[t]
\centering
\subfigure{
\includegraphics[width=5cm]{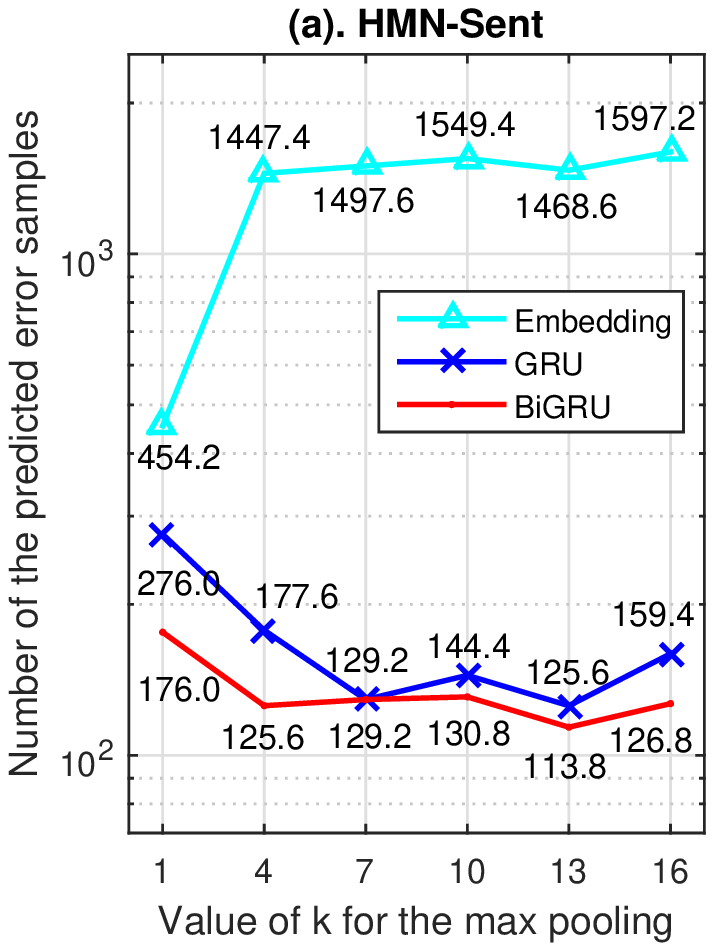}}
\subfigure{
\includegraphics[width=5cm]{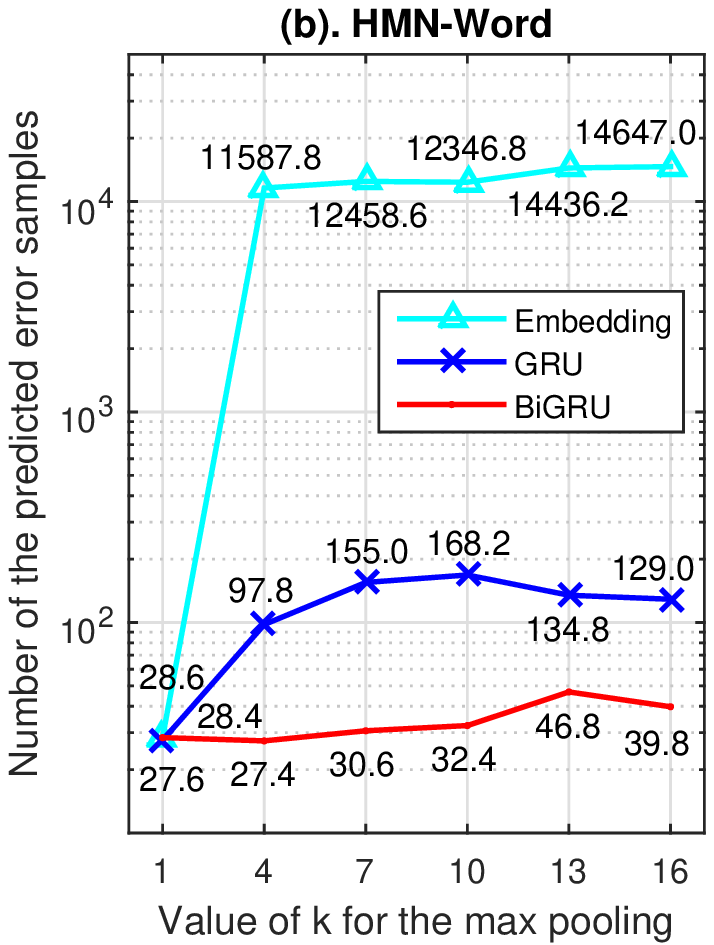}}
\subfigure{
\includegraphics[width=5cm]{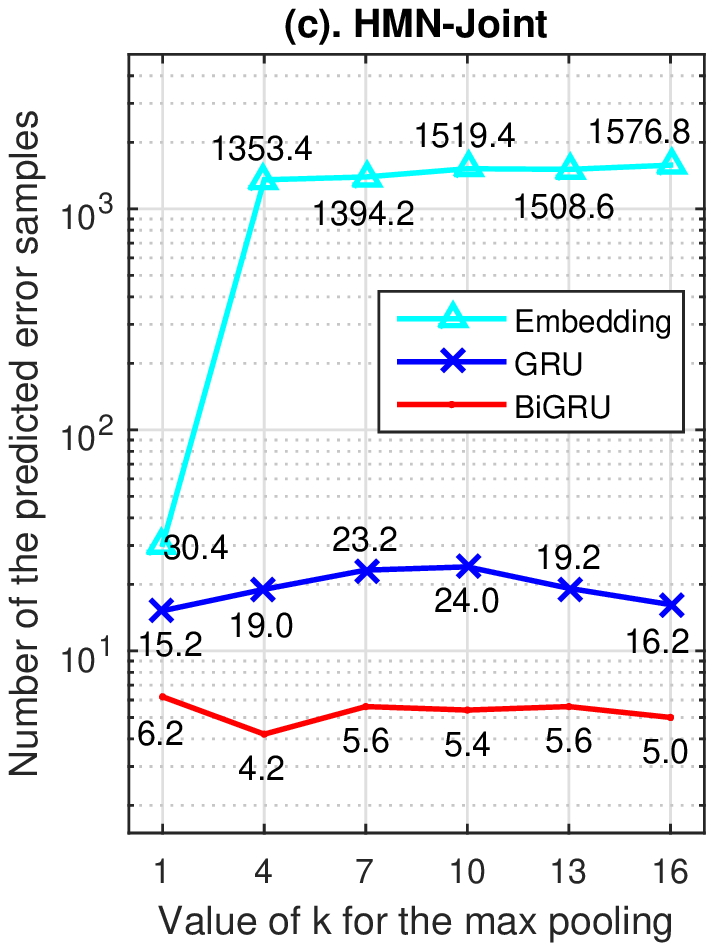}}
\caption{Answer selection evaluations of HMN-Sent, HMN-Word and HMN-Joint with different word-level memory encoding methods (BiGRU, GRU and Embedding) using various values of \(k\) on {\emph{Total}}.}
\label{fig:EffectOfBiGRU}
\end{figure*}

After \(k\)-max pooling for sampling the most relevant sentences, a sophisticated attention mechanism is applied on the selected word-level memory. Two possible reasons make the attention successfully focus on the correct answers: One is that the probe vector \(u_R^{(S)}\) as Eqn.~(\ref{eq:AttentionOfWord}), used in attention mechanism to interact with word-level memory, is generated from the reasoning module and has semantic relevance to the target answers. Another reason is that attention mechanism performs inhibitory effect on our task which can successfully filter the almost useless words, such as ``time'', ``go'' and ``on'' in the example. Besides the above reasons, we also investigate the influence of different word-level memory encoding methods, such as BiGRU (\(\mathop {GRU}\limits^ \to  ({{\bf{C}}^R}{\bar w_t}) + \mathop {GRU}\limits^ \leftarrow  ({{\bf{C}}^R}{\bar w_t})\)), GRU (\(\mathop {GRU}\limits^{} ({{\bf{C}}^R}{\bar w_t})\)) and Embedding (\({{\bf{C}}^R}{\bar w_t}\)), by varying the values of \(k\) for max pooling, and the comparison of answer selection performance are present in Figure~\ref{fig:EffectOfBiGRU}. The results show the expected effect that encoding the context information into word-level memory via BiGRU or GRU can help the attention module distinguish multiple rare and unknown words when we enlarge the value of \(k\) and introduce more unknown words to the attention mechanism.

From Figure~\ref{fig:EffectOfBiGRU}, we further investigate the influence of \(k\) to the answer selection performance. We can see that the performances almost unchanged by using HMN-Joint with BiGRU when we vary the value of \(k\). Considering that the more sentences \(k\)-max pooling samples from sentence-level memory, the more computational complexity the attention mechanism as Eqn.~(\ref{eq:AttentionOfWord}) costs on word-level memory, 4-max pooling used in our experiments is a good trade-off.

\section{Related Works}
\label{sec:RelatedWorks}
Recently, lots of deep learning methods with explicit memory and attention mechanism have shown promising performance in Question Answering (QA) tasks. For example, \newcite{14_yu2015empirical} applied Neural Machine Translation (NMT)~\cite{11_bahdanau2014neural} with sophisticated attention mechanism and Neural Turing Machine (NTM)~\cite{15_graves2014neural} with distributed external memory to solve QA tasks, and \newcite{5_sukhbaatar2015end} designed end-to-end memory networks and introduced multi-hop reasoning component to solve various types of QA task. These representation learning based methods do not rely on any linguistic tools and can be applied to different languages or domains~\cite{16_feng2015applying}. However, most works of these deep learning based methods rarely focus on solving answer selection on unknown word problem. Recently, the unknown word problem has attracted more researchers' attention. \newcite{1_hermann2015teaching} used NLP tools to recognize all the entity and establish co-references to replace all the rare entities by placeholders and trained an attention based model with softmax to predict the placeholder id. \newcite{17_li2016towards} replaced the rare words in a test sentence with similarity in-vocabulary words to solve machine translation task, where the representation of the rare words still can be learned from a large mono-lingual corpus. \newcite{18_gulcehre2016pointing} utilized and extended the attention-based pointing mechanism~\cite{12_vinyals2015pointer} to point the unknown words for machine translation and text summarization. However, the sophisticated attention mechanism is applied on the all word-level representations which may result in high computational complexity, and lots of the fine-grained noise words should be filtered out by reasoning the relevant facts to the query in a high-level semantic space.

\section{Conclusion}
\label{sec:Conclusion}
In this paper, we introduce hierarchical memory networks to solve answer selection problem on unknown words. We first encode the sentences into a sentence-level memory with temporal encoding. Then reasoning module conducts multi-hop interaction on the memory to retrieve the related sentences, and \(k\)-max pooling samples the \(k\) most related sentences. For word-level memory, BiGRU is utilized to encode the words and introduce context into the memory, then a sophisticated attention mechanism is applied on the selected word-level memory to focus the fine-grained words. We conduct answer selection experiments on four synthetic domain dialogue datasets which contain lots of unseen answers. The experimental results show that our hierarchical memory networks can achieve a satisfying performance.

\section*{Acknowledgements}
We thank the anonymous reviewers for their insightful comments, and this work was supported by the Strategic Priority Research Program of the Chinese Academy of Sciences (Grant No. XDB02070005), the National High Technology Research and Development Program of China (863 Program) (Grant No. 2015AA015402) and the National Natural Science Foundation (Grant No. 61602479 and 61403385).
\bibliographystyle{acl}
\bibliography{HMN4QA}

\end{document}